\documentclass[%
 reprint,
 amsmath,amssymb,
 aps,
]{revtex4-1}

\usepackage{graphicx}
\usepackage{dcolumn}
\usepackage{bm}

\newtheorem{definition}{Definition}
\newenvironment{sequation}{\begin{equation}\small} {\end{equation}}
\begin{document}

\preprint{APS/123-QED}

\title{Predicting missing links via significant paths}
\thanks{shimin.cai81@gmail.com}%

\author{Xuzhen Zhu$^1$}
\author{Hui Tian$^1$}
\author{Shimin Cai$^{2*}$}
\author{Tao Zhou$^2$}
\affiliation{%
 $^1$State Key Laboratory of Networking and Switching Technology, Beijing University of Posts and Telecommunications, Beijing, 100876, P.R.China
}%
\affiliation{%
 $^2$Web Sciences Center, University of Electronic Science and Technology of China, Chengdu, 610054, P.R.China
}%

\begin{abstract}
Link prediction plays an important role in understanding the intrinsic evolving mechanisms of networks. With the belief that the likelihood of the existence of a link between two nodes is strongly related to their similarity, many methods have been proposed to calculate node similarity based on node attributes and/or topological structures. Among a large variety of methods that take into account paths connecting the target pair of nodes, most of them neglect the heterogeneity of those paths. Our hypothesis is that a path consisting of small-degree nodes provides a strong evidence of similarity between two ends, accordingly, we propose a so-called \emph{significant path index} in this Letter to leverage intermediate nodes' degrees in similarity calculation. Empirical experiments on twelve disparate real networks demonstrate that the proposed index outperforms the mainstream link prediction baselines.
\begin{description}

\item[PACS numbers]
89.20.Ff, 89.75.Hc, 89.65.-s
\end{description}
\end{abstract}

\maketitle


\section{Introduction}
Plenty of empirical and theoretical studies aim at understanding the intrinsics of evolving networks, from characterizing topological features to revealing network functions~\cite{albert2002statistical,dorogovtsev2002pseudofractal,newman2003structure,boccaletti2006complex,costa2007characterization}. Among fruitful research issues, link prediction is a fundamental challenge~\cite{lu2011link,getoor2005link}. Indeed, the extent to which the network evolution is explicable coincides with our capacity to predict missing links~\cite{wang2012evaluating,zhang2013potential}. In addition, the accurate predictions help recommend friends in online social networks~\cite{scellato2011exploiting,wang2011human}, explore protein-to-protein interactions~\cite{mamitsuka2012mining, cannistraci2013link}, reconstruct airline networks~\cite{guimera2009missing}, boost e-commerce scales~\cite{huang2005link,lu2012recommender}, and so on.

Most conventional methods formalized the task of link prediction in the form of estimating the probability that two disconnected nodes would be linked, which was believed to be positively correlated with the similarity between them~\cite{lu2011link,liben2007link}.
Some researchers extracted node attributes and calculated node similarity in the Euclidean space of attributes, but unfortunately met the difficulty in extraction and sparsity ~\cite{yin2010linkrec,schifanella2010folks}. Instead, mainstream methods took into account topological similarity based on network structures only, and could be classified into three major classes~\cite{lu2011link}. The first class calculated topological similarity with global structural information, such as the \emph{Katz Index} that counts all paths connecting two nodes with shorter paths preferred~\cite{katz1953new}. Such global metrics showed fair performance in prediction but suffered from high computational complexity. The second class, defined on local structures, were much easier to calculate. Typical methods counted the number of common neighbors (CN)~\cite{newman2001clustering}, usually with a penalization of large-degree end nodes (e.g., \emph{Salton Index}~\cite{salton1986introduction}, \emph{S{\o}rensen Index}~\cite{sorensen1948method}, \emph{Hub Promoted Index}~\cite{ravasz2002hierarchical}, \emph{Leicht-Holme-Newman Index}~\cite{leicht2006vertex}, etc.), or of large-degree common neighbors, such as \emph{Adamic-Adar Index} (AA)~\cite{adamic2003friends} and \emph{Resource Allocation Index} (RA)~\cite{zhou2009predicting}. Although successfully reducing the computational expense, local metrics suffered from relatively poor prediction performance. In order to find a nice tradeoff between performance and complexity, the third class of similarity metrics were proposed on quasi-local structures. The \emph{Local Path Index} (LP) ignored long-path terms in Katz Index~\cite{zhou2009predicting,lu2009similarity}, and its bounded version (BLP) relates local paths in an elaborate way~\cite{papadimitriou2012fast}. The \emph{Local Random Walk Index} (LRW) limits a random walker within a local range~\cite{liu2010link}, while the \emph{Superposed Random Walk Index} (SRW) continuously released a random walker at the starting node to emphasize the nodes near the target node~\cite{liu2010link}.

However, most existing methods simply summed paths up and neglected the heterogeneity in paths, i.e., the possibility that different paths, even with the same length, indicate different similarities between two end nodes. Take a problem in collaborative filtering~\cite{goldberg1992using} for example. Since many people has read the popular series of Harry Potter, reading a Harry Potter book does not expose much information of a reader's taste, and therefore finding neighbors for him is sometimes random. The overestimated similarity between such a reader and his dissimilar neighbors attributes to the user-book-user paths containing a large-degree node, the book of Harry Potter. Intuitively speaking, a path containing small-degree nodes indicates a greater similarity. An intermediate node with fewer neighbors is usually more selective in establishing links, and thus has greater similarity with each neighbor. Besides, its small degree sometimes implies a concentrated range of interests. Both reasons lead to a high probability that two of its neighbors are mutually similar. The case of a node with numerous neighbors is the opposite. It is not necessarily that two of its neighbors are mutually similar, e.g., it is probable that a reader with wide interests over various topics have two of his favorite books totally different. With that intuition we believe, two nodes with a small-degree common neighbor are more probable to be similar, which could be extended to a more general form that a path containing small-degree nodes provides an evidence of a greater similarity between two ends.

Inspired by the above discussion, we propose a new index to measure the contribution of a path when calculating the similarity between its two ends, namely the \emph{significant path index}, or \emph{SP} for short. The proposed index incorporates degrees of intermediate nodes into the path length, defining a path is significant if it is short and consists of small-degree intermediate nodes. Such a significant path is believed to reflect a great similarity between its two ends. Empirical results verify that the proposed index significantly improves prediction accuracy in link prediction, compared with five mainstream baselines.

\section{Problem and Metric}

We here consider an undirected network $G(V,E)$, where $V$ and $E$ stand for the sets of nodes and links, respectively. To test the algorithm's accuracy, the set of links, $E$, is randomly divided into two parts: the training set, $E^T$, is treated as known information, while the testing set (also called probe set or validation set in the literature), $E^P$, is used for testing and no information in this set is allowed to be used for prediction. Clearly, $E^T\bigcup E^P = E$ and $E^T \bigcap E^P = \emptyset$. Every link prediction algorithm aims at uncovering links in the testing set.

A standard metric, called the area under the receiver operating characteristic curve (AUC for short)~\cite{hanley1983method} is used to quantify the prediction accuracy. Provided the rank of all non-observed links (i.e., links in $U-E^T$, where $U$ is the universal set) according to their estimated existent likelihoods, the AUC value can be interpreted as the probability that a randomly chosen link in $E^P$ (i.e., a link that indeed exists but does not belong to $E^T$ yet) is ranked higher than a randomly chosen link in $U-E$ (i.e., a non-existent link). If all the likelihood scores are generated from an independent and identical distribution, the AUC value should be about 0.5. Therefore, the degree to which the value exceeds 0.5 indicates how much better the algorithm performs than pure chance.

\section{Baselines}

Under the simplest framework of link prediction~\cite{lu2011link}, each pair of nodes, $v_x$ and $v_y$, is assigned a similarity score $s_{xy}$. All non-observed links are ranked according to their scores, and the links with higher scores are supposed to be of higher existent likelihoods. Five mainstream algorithms (each corresponding to a similarity index) are implemented as baselines for performance comparison, described below.

Common Neighbors Index (CN)~\cite{newman2001clustering} measures the similarity between two nodes with the number of their common neighbors:
\begin{equation}\label{eq:CN}\setlength{\abovedisplayskip}{2pt}\setlength{\belowdisplayskip}{2pt}
s^{CN}_{\mbox{\tiny xy}}=\left|\Gamma(x)\cap\Gamma(y)\right|,
\end{equation}
where $\Gamma(x)$ denotes the set of neighbors of node $v_x$, and $\Gamma(x)\cap\Gamma(y)$ represents the set of common neighbors of node $v_x$ and $v_y$.

Adamic-Adar Index (AA)~\cite{adamic2003friends} expands CN by emphasizing less-connected common neighbors, as:
\begin{equation}\label{eq:AA}\setlength{\abovedisplayskip}{2pt}\setlength{\belowdisplayskip}{2pt}
s^{AA}_{\mbox{\tiny xy}}=\sum_{z\in\Gamma(x)\cap\Gamma(y)}\frac{1}{\log(k_{_z})}.
\end{equation}

Resource-Allocation Index (RA)~\cite{zhou2009predicting,ou2007power} simulates resource transmissions between two nodes, and penalizes common neighbors with large degrees, as:
\begin{equation}\label{eq:RA}\setlength{\abovedisplayskip}{2pt}\setlength{\belowdisplayskip}{2pt}
s^{RA}_{\mbox{\tiny xy}}=\sum_{z\in\Gamma(x)\cap\Gamma(y)}\frac{1}{k_{_z}}.
\end{equation}

Local Path Index (LP)~\cite{zhou2009predicting,lu2009similarity} additionally counts the contribution of local paths with length three, as:
\begin{equation}\label{ep:LP}\setlength{\abovedisplayskip}{2pt}\setlength{\belowdisplayskip}{2pt}
S^{LP}=A^2+\varepsilon A^3,
\end{equation}
where $A$ is the adjacency matrix and $\varepsilon$ is a free parameter.

Bounded Local Path Index (BLP)~\cite{papadimitriou2012fast} bounds local paths with structural coefficients according to path lengths:
\begin{equation}\label{ep:BLP}\setlength{\abovedisplayskip}{2pt}\setlength{\belowdisplayskip}{2pt}
s^{BLP}_{\tiny xy}=\sum_{i=2}^{l_{\max}}\frac{1}{i-1}\cdot\frac{\left|P_i(v_x,v_y)\right|}{\prod_{j=2}^{i}(N-j)},
\end{equation}
where $l_{\max}$ indicates the maximum length under consideration, $P_i(v_x,v_y)$ is the set of all paths connecting $v_x$ and $v_y$ with length $i$, and $\left|P_i(v_x,v_y)\right|$ is the number of these paths.

\section{Significant Paths}
Our basic idea comes from the intuition that paths should not be counted indifferently: a short path consisting of small-degree intermediate nodes provides more evidence of a missing link connecting its two ends. In the proposed \emph{significant path index}, we calculate the similarity between a pair of non-adjacent node by counting paths connecting them while penalizing both the lengths and the intermediate nodes' degrees of those paths.

\begin{definition}
On an undirected unweighted network $G(V,E)$, the significance $\zeta$ of a path $q=\{v_1,v_2,\cdots,v_n\}$ equals to the sum of penalized degree of each intermediate node, as
\begin{equation}
\zeta(q) = \sum_{v_i \in M(q)} k_i^\beta,
\end{equation}
where $k_i$ is the degree of node $v_i$, $M(q)=\{v2_,\cdots,v_{n-1}\}$ is the set of intermediate nodes of path $q$, and $\beta$ is the degree-penalizing parameter that penalizes nodes with large degrees when $\beta<0$. This index has previously been applied to quantify the expected traffic congestion in transportation networks~\cite{yan2006efficient}.
\end{definition}

\begin{definition}
On an undirected unweighted network $G(V,E)$, the significant path index $S_{xy}^{SP}$ of a pair of node $v_x$ and $v_y$ is the sum of significance of every path connecting $v_x$ and $v_y$, as:
\begin{equation}\label{equation:basic-spi}
S_{xy}^{SP} \propto \alpha_1 \sum_{q\in P_2(v_x,v_y)} \zeta(q) + \alpha_2 \sum_{q\in P_3(v_x,v_y)} \zeta(q) + \cdots,
\end{equation}
where $\alpha_{\bullet}$ are the length-penalizing parameters that penalize paths with longer lengths and thus $\alpha_l>\alpha_{l'}$ if $l<l'$.
\end{definition}

Since paths longer than three cost expensive computations but contribute little to predicting links~\cite{lu2009similarity}, we ignore those paths in practice. As only two length-penalizing parameters remain, we simply let $\alpha=\frac{\alpha_2}{\alpha_1}$ and rewrite Eq. (\ref{equation:basic-spi}) as follows:
\begin{sequation}\label{equation:spi}
S_{xy}^{SP} = \sum_{q\in P_2(v_x,v_y)} \sum_{v_i \in M(q)} k_i^\beta + \alpha \sum_{q\in P_3(v_x,v_y)} \sum_{v_i \in M(q)} k_i^\beta .
\end{sequation}

\begin{figure}
\centering
\includegraphics[width=4cm]{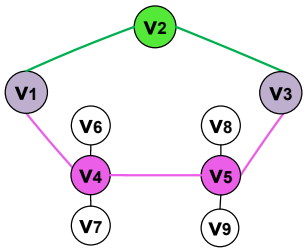}
\caption{An example network to illustrate the significant path index.}
\label{figure:example}
\end{figure}

The two parameters $\alpha$ and $\beta$ penalize path length and node degree respectively, capturing the intuition that the fewer nodes it passes through, the smaller degrees such nodes have, the stronger evidence the two ends of such path is actually connected by a missing link. Figure~\ref{figure:example} shows a simple example to illustrate the significant path index. We calculate the SP index of node $v_1$ and $v_3$, and find a total of $2$ paths connecting them: $q_1=\{v_1,v_2,v_3\}$, $q_2=\{v_1,v_4,v_5,v_3\}$. Comparing those two paths, $v_2$, the only intermediate node of $q_1$, appears more selective and is thus more similar with both $v_1$ and $v_3$, which suggests a larger probability that $v_1$ and $v_3$ are also similar to each other. In opposite, $q_2$ provides less evidence since $q_2$ is longer and its intermediate nodes have larger degrees.
\begin{figure*}[t]
\centering
\includegraphics[width=13.5cm]{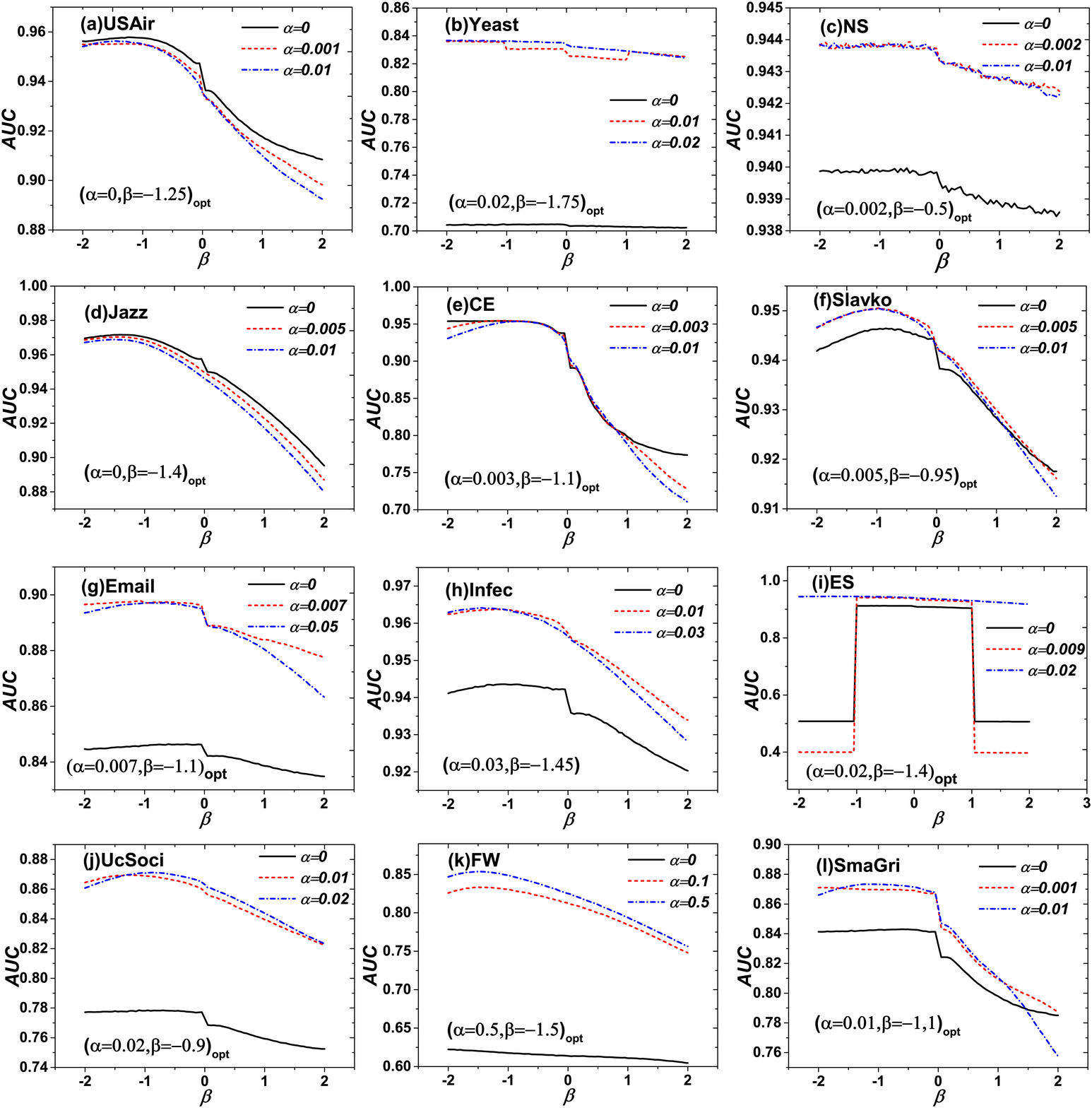}
\caption{The prediction performance of SP index on twelve benchmark networks with different values of $\alpha$ and $\beta$ . For each network, the optimal values of $\alpha$ and $\beta$ are presented inside the corresponding plot. Notice that, for every case, the optimal $\beta$ is smaller than zero, indicating that to penalize the large-degree intermediate nodes.}
\label{fig:AUC}
\end{figure*}
\section{Data}
\begin{table}[t]
\caption{The basic topological features of the twelve benchmark networks. \textit{$\left|\textit{V}\right|$} denotes the number of nodes, $\left|\textit{E}\right|$ is the number of links, $\langle k \rangle$ represents the average degree, $\langle d \rangle$ denotes the average distance, $C$ indicates the clustering coefficient~\cite{watts1998collective}, $r$ indicates the assortativity coefficient~\cite{newman2002assortative}, and $H$ is the degree heterogeneity, defined as $H=\frac{\langle k^2 \rangle}{\langle k\rangle ^2}$.}
\label{tab:data}
\begin{center}
\setlength{\tabcolsep}{4pt}
\begin{tabular}{cccccccc}
\hline\hline
Nets  & $\left|\textit{V}\right|$ & $\left|\textit{E}\right|$ & $\langle k\rangle$ & $\langle d\rangle$ & C & \textit{r} & $H$ \\
\hline
USAir& 332 & 2128 & 12.81 & 2.74 & 0.749 & -0.208 & 3.36 \\
Yeast& 2370 & 10904 & 9.2 & 5.16 & 0.378 & 0.469 & 3.35 \\
NS & 1461 & 2742 & 3.75 & 5.82 & 0.878 & 0.461 & 1.85 \\
Jazz & 198 & 2742 & 27.7 & 2.24 & 0.633 & 0.02 & 1.4 \\
CE & 453 & 2025 & 8.94 & 2.66 & 0.655 & -0.225 & 4.49 \\
Slavko& 334 & 2218 & 13.28 & 3.05 & 0.488 & 0.247 & 1.62 \\
Email & 1133 & 5451 & 9.62 & 3.61 & 0.254 & 0.078 & 1.94\\
Infec & 410 & 2765 & 13.49 & 3.63 & 0.467 & 0.226 & 1.39 \\
ES & 1272 & 6454 & 10.15 & 3.86 & 0.382 & -0.012 & 2.46 \\
UcSoci & 1893 & 13825 & 14.62 & 3.06 & 0.138 & -0.188 & 3.81 \\
FW& 128 & 2075 & 32.42 & 1.78 & 0.334 & -0.112 & 1.24 \\
SmaGri& 1024 & 4916 & 9.6 & 2.98 & 0.349 & -0.193 & 3.95 \\
\hline\hline
\end{tabular}
\end{center}
\end{table}
Experiments are performed on twelve real networks\footnote{Some data sets are freely downloaded from the following academic web sites: http://vlado.fmf.uni-lj.si/pub/networks/data, http://wiki.gephi.org/index.php?title=Datasets, http://lovro.lpt.fri.uni-lj.si/support.jsp, http://konect.uni-koblenz.de/networks/, and http://www.linkprediction.org/index.php/link/resource/data.}. We converted arcs into undirected links and removed loops and multi-links to make them simple networks, keeping their connectivity.
(i) US Air97 (USAir)~\cite{batagelj1998pajek}: the network of the US air transportation system.
(ii) Yeast PPI (Yeast)~\cite{Yeast}: the protein-protein interaction network of yeast.
(iii) NetScience (NS)~\cite{newman2006finding}: the network of coauthorships between scientists publishing on the topic of networks.
(iv) Jazz~\cite{gleiser2003community}: the network of Jazz musicians.
(v) C.elegans (CE)~\cite{watts1998collective}: the neural network of the nematode worm C.elegans.
(vi) Slavko~\cite{blagus2012self}: the Facebook friendship network of Slavko $\check{Z}$itnik.
(vii) Email network (Email)~\cite{guimera2003self}: the email communication network of University Rovira i Virgili (URV) in
Tarragona, Spain.
(viii) Infectious (Infec)~\cite{isella2011s}: the face-to-face contact network of people during the exhibition ``Infectious: Stay Away" in 2009 at the Science Gallery in Dublin.
(ix) EuroSiS web mapping study (ES)~\cite{van2004mapping}: mapping interactions between Science in Society actors on the Web of 12 European countries.
(x) UC Irvine messages social network (UcSoci)~\cite{opsahl2009clustering}: the communication network according to the sent messages between users of an online community of students from the University of California, Irvine.
(xi) Food Web of Florida ecosystem (FW)~\cite{bascompte2004food}: the relationship of carbon exchanges in the cypress wetlands of South Florida during the wet season.
(xii) Small \& Griffith and Descendants (SmaGri)~\cite{hummon1989connectivity}: the network composed of citations to Small \& Griffith and Descendants.
Table~\ref{tab:data} reports the basic topological features of these networks.

Each original data set is randomly divided into a training set $E^T$ containing $80\%$ links, and a testing set $E^P$ containing $20\%$ links, while the connectivity in $E^T$ is guaranteed.

\begin{table*}[t]
\centering
\caption{Prediction accuracy measured by AUC values on the twelve benchmark networks. Each data point is an average over 10 independent realizations, each of which corresponds to a random 80\%-20\% division of training set and testing set. All the present results corresponding to the optimal cases by tuning the parameters if any. Numbers in brackets stand for the standard deviations.}
\label{table:AUC}
\setlength{\tabcolsep}{6pt}
\renewcommand{\arraystretch}{1.2}
\begin{tabular}{ccccccc}
\hline\hline
AUC & CN & AA & RA & LP & BLP & SP \\
\hline
USAir & 0.938(0.0064) & 0.950(0.0072) & 0.956(0.0075) & 0.938(0.0073) & 0.931(0.0100) &\textbf{0.960(0.0132)}\\
Yeast & 0.703(0.0053) & 0.705(0.0051) & 0.705(0.0052) & 0.780(0.0068) & 0.836(0.0089) & \textbf{0.847(0.0086)}\\
NS	& 0.940(0.0114) & 0.940(0.0114) & 0.940(0.0114) & 0.940(0.0118) & 0.943(0.0095) & \textbf{0.944(0.0094)}\\
Jazz & 0.954(0.0054) & 0.961(0.0047) & 0.970(0.0046) & 0.954(0.0053) & 0.951(0.0056) & \textbf{0.972(0.0044)}\\
CE & 0.914(0.0119) & 0.948(0.0102) & 0.954(0.0101) & 0.914(0.0112) & 0.911(0.0076) & \textbf{0.957(0.0089)}\\
Slavko & 0.941(0.0098)& 0.945(0.0099) & 0.946(0.0100) & 0.944(0.0101) & 0.943(0.0104) & \textbf{0.951(0.0096)}\\
Email & 0.844(0.0071) & 0.846(0.0071)& 0.846(0.0070)& 0.893(0.0064)& \textbf{0.902(0.0054)}& 0.899(0.0077)\\
Infec & 0.939(0.0096)& 0.943(0.0094)& 0.944(0.0093)& 0.954(0.0124)& 0.958(0.0065)& \textbf{0.964(0.0062)}\\
ES	& 0.910(0.0059)& 0.912(0.0061)& 0.912(0.0061)& 0.936(0.0073)& 0.938(0.0051)& \textbf{0.945(0.0056)}\\
UcSoci	& 0.773(0.0066)& 0.778(0.0068)& 0.778(0.0068)& 0.838(0.0051)& 0.870(0.0059)& \textbf{0.871(0.0073)}\\
FW & 0.612(0.0162)& 0.615(0.0158)& 0.620(0.0146)& 0.800(0.0121)& 0.641(0.0051)& \textbf{0.873(0.0142)}\\
SmaGri & 0.833(0.0074) &0.843(0.0073) &0.843(0.0076) &0.857(0.0079) &\textbf{0.875(0.0076)} &0.874(0.0103)\\
\hline\hline
\end{tabular}
\end{table*}

\section{Results}

We firstly report the performance of SP with $\beta$ varying in the range $[-2, 2]$ and $\alpha$ from $0$ to $1.0$ as suggested in~\cite{lu2009similarity}, measured with average AUC under $10$ independent runs with different random divisions of training sets and testing set. As shown in Figure~\ref{fig:AUC}, the performance varies with $\beta$ continuously in most cases. For every data set, a single peak is observed when $\beta < 0$. It demonstrates the intuition we claimed that the intermediate nodes with smaller degrees provide stronger evidence of the existence of a missing link than nodes with larger degrees. Specifically, in most data sets a sudden decrease around $\beta=0$ is observed, at which the index changes from penalizing paths consisting of large-degree intermediate nodes ($\beta<0$) to penalizing paths consisting of small-degree intermediate nodes ($\beta>0$). Since large-degree nodes usually lie in many paths (as suggested by the strong correlation between degree and betweenness, see for example~\cite{guimera2005worldwide}), in the case of $\beta>0$, many paths with large-degree nodes are wrongly assigned high contributions and thus a large proportion of nonexistent links will be put in the top positions, eventually leading to a large decline in algorithm's performance.

The performance seems not sensitive to $\alpha$ as long as $\alpha >0$. In Figure~\ref{fig:AUC}, besides curves when $\alpha$ takes typical positive values such as $0.001$ and $0.01$, we intentionally provide the curve when $\alpha=0$, meaning deletion of long paths' contributions. In most data sets, except for \textit{USAir} and \textit{Jazz}, there is a distinctive deviation between curves with $\alpha=0$ and $\alpha>0$. The AUC performance curve with $\alpha=0$ is obviously beneath the contrasts. 
Accordingly, SP is particularly useful for the networks with many long paths. The phenomenon indeed confirms that combination with long paths is very necessary.

To demonstrate the prediction ability, we report the performance of SP index with the optimal $(\alpha,\beta)$ value on $12$ data sets respectively. Table~\ref{table:AUC} reports the average AUC values of SP and baselines. SP achieves the best performance in $10$ out of $12$ data sets, and is the runner-up in the remaining two (see boldface in Table~\ref{table:AUC}). Notice that those data sets represent different kinds of networks with heterogeneous topological features (see Table 1) and disparate organization principles, the comparison highlights that SP works well consistently on various situations.

Analyzing the difference in performance between SP and baselines, we realize that it is the penalization on large-degree nodes and consideration of long paths that explain the difference. CN indifferently counts the number of common neighbors on 2-hop paths, ignoring node degrees and long paths, resulting in its worst performance in most data sets. AA and RA extend CN by similarly penalizing large-degree nodes, and not surprisingly obtain better performance than CN. However, the fixed form of penalization limits their flexibility to adjust for disparate networks. Besides, ignoring long paths prevents them from better predictions in networks fulfilled with long paths such as \textit{Yeast}. In contrast, LP and BLP take long paths into account and thus outperform on those two networks. However, lacking consideration of large-degree nodes, LP and BLP meets difficulty in accurately predict missing links in networks with plenty of large-degree nodes, such as \textit{USAir}.

\section{Conclusion and Discussion}

We start from the intuition that a path containing small-degree nodes indicates a greater similarity, and thus provide more evidence of a missing link between its two ends. A \emph{Significant Path index} (SP) is proposed to formalize that intuition into a similarity metric, summing up the discounted degrees of all intermediate nodes over all local paths connecting two non-adjacent nodes. Empirical results demonstrate that, with a negative parameter $\beta$ that penalizes large-degree nodes, the proposed SP outperforms five representative mainstream baselines, evaluated with AUC as binary classifiers. A sudden decrease in prediction performance is observed when $\beta$ changes from penalizing to encouraging large-degree nodes.

Looking at the parameters $\alpha$ and $\beta$, we find that instead of a fixed value of penalizing parameter, i.e., $\beta=-1$ in RA, the SP allows flexible value of penalizing parameter $\beta$, resulting in its performance among the best in most data sets. The contribution of long paths are weak but not negligible. In networks where long paths play a more important role, ignoring long paths is unwise and leads to poor performance, suggested by the performance of RA in \textit{Yeast}, meanwhile an inappropriately large $\alpha$ also depresses the performance.

The SP starts a broad space for investigation on link prediction and facilitates new promising applications in the future, such as infrastructure constructions planning like traffic transportation, researches on the reactions between organisms in biological experiments, evolutions of people's relations in social activities, controls of propagation of disease, protections of ecological system network, and so on.

\begin{acknowledgments}
This work was supported by National Natural Science Foundation of China (No.61302077), National Major Science and Technology Special Project of China (No.2014AA01A706), Funds for Creative Research Groups of China (No.61121001) and EU FP7 Project EVANS (No.2010-269323). TZ acknowledges the Program for New Century Excellent Talents in University under Grant No. NCET-11-0070 and the Special Project of Sichuan Youth Science and Technology Innovation Research Team (Grant No. 2013TD0006).
\end{acknowledgments}

\bibliography{apssamp}

\begin{thebibliography}{49}%
\makeatletter
\providecommand \@ifxundefined [1]{%
 \@ifx{#1\undefined}
}%
\providecommand \@ifnum [1]{%
 \ifnum #1\expandafter \@firstoftwo
 \else \expandafter \@secondoftwo
 \fi
}%
\providecommand \@ifx [1]{%
 \ifx #1\expandafter \@firstoftwo
 \else \expandafter \@secondoftwo
 \fi
}%
\providecommand \natexlab [1]{#1}%
\providecommand \enquote  [1]{``#1''}%
\providecommand \bibnamefont  [1]{#1}%
\providecommand \bibfnamefont [1]{#1}%
\providecommand \citenamefont [1]{#1}%
\providecommand \href@noop [0]{\@secondoftwo}%
\providecommand \href [0]{\begingroup \@sanitize@url \@href}%
\providecommand \@href[1]{\@@startlink{#1}\@@href}%
\providecommand \@@href[1]{\endgroup#1\@@endlink}%
\providecommand \@sanitize@url [0]{\catcode `\\12\catcode `\$12\catcode
  `\&12\catcode `\#12\catcode `\^12\catcode `\_12\catcode `\%12\relax}%
\providecommand \@@startlink[1]{}%
\providecommand \@@endlink[0]{}%
\providecommand \url  [0]{\begingroup\@sanitize@url \@url }%
\providecommand \@url [1]{\endgroup\@href {#1}{\urlprefix }}%
\providecommand \urlprefix  [0]{URL }%
\providecommand \Eprint [0]{\href }%
\providecommand \doibase [0]{http://dx.doi.org/}%
\providecommand \selectlanguage [0]{\@gobble}%
\providecommand \bibinfo  [0]{\@secondoftwo}%
\providecommand \bibfield  [0]{\@secondoftwo}%
\providecommand \translation [1]{[#1]}%
\providecommand \BibitemOpen [0]{}%
\providecommand \bibitemStop [0]{}%
\providecommand \bibitemNoStop [0]{.\EOS\space}%
\providecommand \EOS [0]{\spacefactor3000\relax}%
\providecommand \BibitemShut  [1]{\csname bibitem#1\endcsname}%
\let\auto@bib@innerbib\@empty
\bibitem [{\citenamefont {Albert}\ and\ \citenamefont
  {Barab{\'a}si}(2002)}]{albert2002statistical}%
  \BibitemOpen
  \bibfield  {author} {\bibinfo {author} {\bibfnamefont {R.}~\bibnamefont
  {Albert}}\ and\ \bibinfo {author} {\bibfnamefont {A.-L.}\ \bibnamefont
  {Barab{\'a}si}},\ }\href@noop {} {\bibfield  {journal} {\bibinfo  {journal}
  {Rev. Mod. Phys.}\ }\textbf {\bibinfo {volume} {74}},\ \bibinfo {pages} {47}
  (\bibinfo {year} {2002})}\BibitemShut {NoStop}%
\bibitem [{\citenamefont {Dorogovtsev}\ \emph {et~al.}(2002)\citenamefont
  {Dorogovtsev}, \citenamefont {Goltsev},\ and\ \citenamefont
  {Mendes}}]{dorogovtsev2002pseudofractal}%
  \BibitemOpen
  \bibfield  {author} {\bibinfo {author} {\bibfnamefont {S.~N.}\ \bibnamefont
  {Dorogovtsev}}, \bibinfo {author} {\bibfnamefont {A.}~\bibnamefont
  {Goltsev}}, \ and\ \bibinfo {author} {\bibfnamefont {J.~F.~F.}\ \bibnamefont
  {Mendes}},\ }\href@noop {} {\bibfield  {journal} {\bibinfo  {journal} {Phys.
  Rev. E}\ }\textbf {\bibinfo {volume} {65}},\ \bibinfo {pages} {066122}
  (\bibinfo {year} {2002})}\BibitemShut {NoStop}%
\bibitem [{\citenamefont {Newman}(2003)}]{newman2003structure}%
  \BibitemOpen
  \bibfield  {author} {\bibinfo {author} {\bibfnamefont {M.~E.~J.}\
  \bibnamefont {Newman}},\ }\href@noop {} {\bibfield  {journal} {\bibinfo
  {journal} {SIAM Rev.}\ }\textbf {\bibinfo {volume} {45}},\ \bibinfo {pages}
  {167} (\bibinfo {year} {2003})}\BibitemShut {NoStop}%
\bibitem [{\citenamefont {Boccaletti}\ \emph {et~al.}(2006)\citenamefont
  {Boccaletti}, \citenamefont {Latora}, \citenamefont {Moreno}, \citenamefont
  {Chavez},\ and\ \citenamefont {Hwang}}]{boccaletti2006complex}%
  \BibitemOpen
  \bibfield  {author} {\bibinfo {author} {\bibfnamefont {S.}~\bibnamefont
  {Boccaletti}}, \bibinfo {author} {\bibfnamefont {V.}~\bibnamefont {Latora}},
  \bibinfo {author} {\bibfnamefont {Y.}~\bibnamefont {Moreno}}, \bibinfo
  {author} {\bibfnamefont {M.}~\bibnamefont {Chavez}}, \ and\ \bibinfo {author}
  {\bibfnamefont {D.-U.}\ \bibnamefont {Hwang}},\ }\href@noop {} {\bibfield
  {journal} {\bibinfo  {journal} {Phys. Rep.}\ }\textbf {\bibinfo {volume}
  {424}},\ \bibinfo {pages} {175} (\bibinfo {year} {2006})}\BibitemShut
  {NoStop}%
\bibitem [{\citenamefont {Costa}\ \emph {et~al.}(2007)\citenamefont {Costa},
  \citenamefont {Rodrigues}, \citenamefont {Travieso},\ and\ \citenamefont
  {Villas~Boas}}]{costa2007characterization}%
  \BibitemOpen
  \bibfield  {author} {\bibinfo {author} {\bibfnamefont {L.~D.~F.}\
  \bibnamefont {Costa}}, \bibinfo {author} {\bibfnamefont {F.~A.}\ \bibnamefont
  {Rodrigues}}, \bibinfo {author} {\bibfnamefont {G.}~\bibnamefont {Travieso}},
  \ and\ \bibinfo {author} {\bibfnamefont {P.}~\bibnamefont {Villas~Boas}},\
  }\href@noop {} {\bibfield  {journal} {\bibinfo  {journal} {Adv. Phys.}\
  }\textbf {\bibinfo {volume} {56}},\ \bibinfo {pages} {167} (\bibinfo {year}
  {2007})}\BibitemShut {NoStop}%
\bibitem [{\citenamefont {L{\"u}}\ and\ \citenamefont
  {Zhou}(2011)}]{lu2011link}%
  \BibitemOpen
  \bibfield  {author} {\bibinfo {author} {\bibfnamefont {L.}~\bibnamefont
  {L{\"u}}}\ and\ \bibinfo {author} {\bibfnamefont {T.}~\bibnamefont {Zhou}},\
  }\href@noop {} {\bibfield  {journal} {\bibinfo  {journal} {Physica A}\
  }\textbf {\bibinfo {volume} {390}},\ \bibinfo {pages} {1150} (\bibinfo {year}
  {2011})}\BibitemShut {NoStop}%
\bibitem [{\citenamefont {Getoor}\ and\ \citenamefont
  {Diehl}(2005)}]{getoor2005link}%
  \BibitemOpen
  \bibfield  {author} {\bibinfo {author} {\bibfnamefont {L.}~\bibnamefont
  {Getoor}}\ and\ \bibinfo {author} {\bibfnamefont {C.~P.}\ \bibnamefont
  {Diehl}},\ }\href@noop {} {\bibfield  {journal} {\bibinfo  {journal} {ACM
  SIGKDD Explor. Newsl.}\ }\textbf {\bibinfo {volume} {7}},\ \bibinfo {pages}
  {3} (\bibinfo {year} {2005})}\BibitemShut {NoStop}%
\bibitem [{\citenamefont {Wang}\ \emph {et~al.}(2012)\citenamefont {Wang},
  \citenamefont {Zhang},\ and\ \citenamefont {Zhou}}]{wang2012evaluating}%
  \BibitemOpen
  \bibfield  {author} {\bibinfo {author} {\bibfnamefont {W.-Q.}\ \bibnamefont
  {Wang}}, \bibinfo {author} {\bibfnamefont {Q.-M.}\ \bibnamefont {Zhang}}, \
  and\ \bibinfo {author} {\bibfnamefont {T.}~\bibnamefont {Zhou}},\ }\href@noop
  {} {\bibfield  {journal} {\bibinfo  {journal} {EPL}\ }\textbf {\bibinfo
  {volume} {98}},\ \bibinfo {pages} {28004} (\bibinfo {year}
  {2012})}\BibitemShut {NoStop}%
\bibitem [{\citenamefont {Zhang}\ \emph {et~al.}(2013)\citenamefont {Zhang},
  \citenamefont {L{\"u}}, \citenamefont {Wang},\ and\ \citenamefont
  {Zhou}}]{zhang2013potential}%
  \BibitemOpen
  \bibfield  {author} {\bibinfo {author} {\bibfnamefont {Q.-M.}\ \bibnamefont
  {Zhang}}, \bibinfo {author} {\bibfnamefont {L.}~\bibnamefont {L{\"u}}},
  \bibinfo {author} {\bibfnamefont {W.-Q.}\ \bibnamefont {Wang}}, \ and\
  \bibinfo {author} {\bibfnamefont {T.}~\bibnamefont {Zhou}},\ }\href@noop {}
  {\bibfield  {journal} {\bibinfo  {journal} {PLoS ONE}\ }\textbf {\bibinfo
  {volume} {8}},\ \bibinfo {pages} {e55437} (\bibinfo {year}
  {2013})}\BibitemShut {NoStop}%
\bibitem [{\citenamefont {Scellato}\ \emph {et~al.}(2011)\citenamefont
  {Scellato}, \citenamefont {Noulas},\ and\ \citenamefont
  {Mascolo}}]{scellato2011exploiting}%
  \BibitemOpen
  \bibfield  {author} {\bibinfo {author} {\bibfnamefont {S.}~\bibnamefont
  {Scellato}}, \bibinfo {author} {\bibfnamefont {A.}~\bibnamefont {Noulas}}, \
  and\ \bibinfo {author} {\bibfnamefont {C.}~\bibnamefont {Mascolo}},\ }in\
  \href@noop {} {\emph {\bibinfo {booktitle} {the 17th ACM SIGKDD international
  conference on Knowledge discovery and data mining}}}\ (\bibinfo
  {organization} {ACM},\ \bibinfo {year} {2011})\ pp.\ \bibinfo {pages}
  {1046--1054}\BibitemShut {NoStop}%
\bibitem [{\citenamefont {Wang}\ \emph {et~al.}(2011)\citenamefont {Wang},
  \citenamefont {Pedreschi}, \citenamefont {Song}, \citenamefont {Giannotti},\
  and\ \citenamefont {Barabasi}}]{wang2011human}%
  \BibitemOpen
  \bibfield  {author} {\bibinfo {author} {\bibfnamefont {D.}~\bibnamefont
  {Wang}}, \bibinfo {author} {\bibfnamefont {D.}~\bibnamefont {Pedreschi}},
  \bibinfo {author} {\bibfnamefont {C.}~\bibnamefont {Song}}, \bibinfo {author}
  {\bibfnamefont {F.}~\bibnamefont {Giannotti}}, \ and\ \bibinfo {author}
  {\bibfnamefont {A.-L.}\ \bibnamefont {Barabasi}},\ }in\ \href@noop {} {\emph
  {\bibinfo {booktitle} {Proceedings of the 17th ACM SIGKDD international
  conference on Knowledge discovery and data mining}}}\ (\bibinfo
  {organization} {ACM},\ \bibinfo {year} {2011})\ pp.\ \bibinfo {pages}
  {1100--1108}\BibitemShut {NoStop}%
\bibitem [{\citenamefont {Mamitsuka}(2012)}]{mamitsuka2012mining}%
  \BibitemOpen
  \bibfield  {author} {\bibinfo {author} {\bibfnamefont {H.}~\bibnamefont
  {Mamitsuka}},\ }\href@noop {} {\bibfield  {journal} {\bibinfo  {journal}
  {Data Min. Knowl. Disc.}\ }\textbf {\bibinfo {volume} {2}},\ \bibinfo {pages}
  {400} (\bibinfo {year} {2012})}\BibitemShut {NoStop}%
\bibitem [{\citenamefont {Cannistraci}\ \emph {et~al.}(2013)\citenamefont
  {Cannistraci}, \citenamefont {Alanis-Lobato},\ and\ \citenamefont
  {Ravasi}}]{cannistraci2013link}%
  \BibitemOpen
  \bibfield  {author} {\bibinfo {author} {\bibfnamefont {C.~V.}\ \bibnamefont
  {Cannistraci}}, \bibinfo {author} {\bibfnamefont {G.}~\bibnamefont
  {Alanis-Lobato}}, \ and\ \bibinfo {author} {\bibfnamefont {T.}~\bibnamefont
  {Ravasi}},\ }\href@noop {} {\bibfield  {journal} {\bibinfo  {journal} {Sci.
  Rep.}\ }\textbf {\bibinfo {volume} {3}} (\bibinfo {year} {2013})}\BibitemShut
  {NoStop}%
\bibitem [{\citenamefont {Guimer{\`a}}\ and\ \citenamefont
  {Sales-Pardo}(2009)}]{guimera2009missing}%
  \BibitemOpen
  \bibfield  {author} {\bibinfo {author} {\bibfnamefont {R.}~\bibnamefont
  {Guimer{\`a}}}\ and\ \bibinfo {author} {\bibfnamefont {M.}~\bibnamefont
  {Sales-Pardo}},\ }\href@noop {} {\bibfield  {journal} {\bibinfo  {journal}
  {Proc. Natl. Acad. Sci. USA}\ }\textbf {\bibinfo {volume} {106}},\ \bibinfo
  {pages} {22073} (\bibinfo {year} {2009})}\BibitemShut {NoStop}%
\bibitem [{\citenamefont {Huang}\ \emph {et~al.}(2005)\citenamefont {Huang},
  \citenamefont {Li},\ and\ \citenamefont {Chen}}]{huang2005link}%
  \BibitemOpen
  \bibfield  {author} {\bibinfo {author} {\bibfnamefont {Z.}~\bibnamefont
  {Huang}}, \bibinfo {author} {\bibfnamefont {X.}~\bibnamefont {Li}}, \ and\
  \bibinfo {author} {\bibfnamefont {H.}~\bibnamefont {Chen}},\ }in\ \href@noop
  {} {\emph {\bibinfo {booktitle} {the 5th ACM/IEEE-CS Joint Conference on
  Digital Libraries}}}\ (\bibinfo {organization} {ACM},\ \bibinfo {year}
  {2005})\ pp.\ \bibinfo {pages} {141--142}\BibitemShut {NoStop}%
\bibitem [{\citenamefont {L{\"u}}\ \emph {et~al.}(2012)\citenamefont {L{\"u}},
  \citenamefont {Medo}, \citenamefont {Yeung}, \citenamefont {Zhang},
  \citenamefont {Zhang},\ and\ \citenamefont {Zhou}}]{lu2012recommender}%
  \BibitemOpen
  \bibfield  {author} {\bibinfo {author} {\bibfnamefont {L.}~\bibnamefont
  {L{\"u}}}, \bibinfo {author} {\bibfnamefont {M.}~\bibnamefont {Medo}},
  \bibinfo {author} {\bibfnamefont {C.~H.}\ \bibnamefont {Yeung}}, \bibinfo
  {author} {\bibfnamefont {Y.-C.}\ \bibnamefont {Zhang}}, \bibinfo {author}
  {\bibfnamefont {Z.-K.}\ \bibnamefont {Zhang}}, \ and\ \bibinfo {author}
  {\bibfnamefont {T.}~\bibnamefont {Zhou}},\ }\href@noop {} {\bibfield
  {journal} {\bibinfo  {journal} {Phys. Rep.}\ }\textbf {\bibinfo {volume}
  {519}},\ \bibinfo {pages} {1} (\bibinfo {year} {2012})}\BibitemShut {NoStop}%
\bibitem [{\citenamefont {Liben-Nowell}\ and\ \citenamefont
  {Kleinberg}(2007)}]{liben2007link}%
  \BibitemOpen
  \bibfield  {author} {\bibinfo {author} {\bibfnamefont {D.}~\bibnamefont
  {Liben-Nowell}}\ and\ \bibinfo {author} {\bibfnamefont {J.}~\bibnamefont
  {Kleinberg}},\ }\href@noop {} {\bibfield  {journal} {\bibinfo  {journal} {J.
  Am. Soc. Inf. Sci. Technol.}\ }\textbf {\bibinfo {volume} {58}},\ \bibinfo
  {pages} {1019} (\bibinfo {year} {2007})}\BibitemShut {NoStop}%
\bibitem [{\citenamefont {Yin}\ \emph {et~al.}(2010)\citenamefont {Yin},
  \citenamefont {Gupta}, \citenamefont {Weninger},\ and\ \citenamefont
  {Han}}]{yin2010linkrec}%
  \BibitemOpen
  \bibfield  {author} {\bibinfo {author} {\bibfnamefont {Z.}~\bibnamefont
  {Yin}}, \bibinfo {author} {\bibfnamefont {M.}~\bibnamefont {Gupta}}, \bibinfo
  {author} {\bibfnamefont {T.}~\bibnamefont {Weninger}}, \ and\ \bibinfo
  {author} {\bibfnamefont {J.}~\bibnamefont {Han}},\ }in\ \href@noop {} {\emph
  {\bibinfo {booktitle} {the 19th international conference on World wide
  web}}}\ (\bibinfo {organization} {ACM},\ \bibinfo {year} {2010})\ pp.\
  \bibinfo {pages} {1211--1212}\BibitemShut {NoStop}%
\bibitem [{\citenamefont {Schifanella}\ \emph {et~al.}(2010)\citenamefont
  {Schifanella}, \citenamefont {Barrat}, \citenamefont {Cattuto}, \citenamefont
  {Markines},\ and\ \citenamefont {Menczer}}]{schifanella2010folks}%
  \BibitemOpen
  \bibfield  {author} {\bibinfo {author} {\bibfnamefont {R.}~\bibnamefont
  {Schifanella}}, \bibinfo {author} {\bibfnamefont {A.}~\bibnamefont {Barrat}},
  \bibinfo {author} {\bibfnamefont {C.}~\bibnamefont {Cattuto}}, \bibinfo
  {author} {\bibfnamefont {B.}~\bibnamefont {Markines}}, \ and\ \bibinfo
  {author} {\bibfnamefont {F.}~\bibnamefont {Menczer}},\ }in\ \href@noop {}
  {\emph {\bibinfo {booktitle} {the third ACM international conference on Web
  search and data mining}}}\ (\bibinfo {organization} {ACM},\ \bibinfo {year}
  {2010})\ pp.\ \bibinfo {pages} {271--280}\BibitemShut {NoStop}%
\bibitem [{\citenamefont {Katz}(1953)}]{katz1953new}%
  \BibitemOpen
  \bibfield  {author} {\bibinfo {author} {\bibfnamefont {L.}~\bibnamefont
  {Katz}},\ }\href@noop {} {\bibfield  {journal} {\bibinfo  {journal}
  {Psychometrika}\ }\textbf {\bibinfo {volume} {18}},\ \bibinfo {pages} {39}
  (\bibinfo {year} {1953})}\BibitemShut {NoStop}%
\bibitem [{\citenamefont {Newman}(2001)}]{newman2001clustering}%
  \BibitemOpen
  \bibfield  {author} {\bibinfo {author} {\bibfnamefont {M.~E.~J.}\
  \bibnamefont {Newman}},\ }\href@noop {} {\bibfield  {journal} {\bibinfo
  {journal} {Phys. Rev. E}\ }\textbf {\bibinfo {volume} {64}},\ \bibinfo
  {pages} {025102} (\bibinfo {year} {2001})}\BibitemShut {NoStop}%
\bibitem [{\citenamefont {Salton}\ and\ \citenamefont
  {McGill}(1986)}]{salton1986introduction}%
  \BibitemOpen
  \bibfield  {author} {\bibinfo {author} {\bibfnamefont {G.}~\bibnamefont
  {Salton}}\ and\ \bibinfo {author} {\bibfnamefont {M.~J.}\ \bibnamefont
  {McGill}},\ }\href@noop {} {\emph {\bibinfo {title} {Introduction to Modern
  Information Retrieval}}}\ (\bibinfo  {publisher} {McGraw-Hill, Inc.},\
  \bibinfo {year} {1986})\BibitemShut {NoStop}%
\bibitem [{\citenamefont {S{\o}rensen}(1948)}]{sorensen1948method}%
  \BibitemOpen
  \bibfield  {author} {\bibinfo {author} {\bibfnamefont {T.}~\bibnamefont
  {S{\o}rensen}},\ }\href@noop {} {\bibfield  {journal} {\bibinfo  {journal}
  {Biol. Skr.}\ }\textbf {\bibinfo {volume} {5}},\ \bibinfo {pages} {1}
  (\bibinfo {year} {1948})}\BibitemShut {NoStop}%
\bibitem [{\citenamefont {Ravasz}\ \emph {et~al.}(2002)\citenamefont {Ravasz},
  \citenamefont {Somera}, \citenamefont {Mongru}, \citenamefont {Oltvai},\ and\
  \citenamefont {Barab{\'a}si}}]{ravasz2002hierarchical}%
  \BibitemOpen
  \bibfield  {author} {\bibinfo {author} {\bibfnamefont {E.}~\bibnamefont
  {Ravasz}}, \bibinfo {author} {\bibfnamefont {A.~L.}\ \bibnamefont {Somera}},
  \bibinfo {author} {\bibfnamefont {D.~A.}\ \bibnamefont {Mongru}}, \bibinfo
  {author} {\bibfnamefont {Z.~N.}\ \bibnamefont {Oltvai}}, \ and\ \bibinfo
  {author} {\bibfnamefont {A.-L.}\ \bibnamefont {Barab{\'a}si}},\ }\href@noop
  {} {\bibfield  {journal} {\bibinfo  {journal} {Science}\ }\textbf {\bibinfo
  {volume} {297}},\ \bibinfo {pages} {1551} (\bibinfo {year}
  {2002})}\BibitemShut {NoStop}%
\bibitem [{\citenamefont {Leicht}\ \emph {et~al.}(2006)\citenamefont {Leicht},
  \citenamefont {Holme},\ and\ \citenamefont {Newman}}]{leicht2006vertex}%
  \BibitemOpen
  \bibfield  {author} {\bibinfo {author} {\bibfnamefont {E.}~\bibnamefont
  {Leicht}}, \bibinfo {author} {\bibfnamefont {P.}~\bibnamefont {Holme}}, \
  and\ \bibinfo {author} {\bibfnamefont {M.~E.~J.}\ \bibnamefont {Newman}},\
  }\href@noop {} {\bibfield  {journal} {\bibinfo  {journal} {Phys. Rev. E}\
  }\textbf {\bibinfo {volume} {73}},\ \bibinfo {pages} {026120} (\bibinfo
  {year} {2006})}\BibitemShut {NoStop}%
\bibitem [{\citenamefont {Adamic}\ and\ \citenamefont
  {Adar}(2003)}]{adamic2003friends}%
  \BibitemOpen
  \bibfield  {author} {\bibinfo {author} {\bibfnamefont {L.~A.}\ \bibnamefont
  {Adamic}}\ and\ \bibinfo {author} {\bibfnamefont {E.}~\bibnamefont {Adar}},\
  }\href@noop {} {\bibfield  {journal} {\bibinfo  {journal} {Soc. Netw.}\
  }\textbf {\bibinfo {volume} {25}},\ \bibinfo {pages} {211} (\bibinfo {year}
  {2003})}\BibitemShut {NoStop}%
\bibitem [{\citenamefont {Zhou}\ \emph {et~al.}(2009)\citenamefont {Zhou},
  \citenamefont {L{\"u}},\ and\ \citenamefont {Zhang}}]{zhou2009predicting}%
  \BibitemOpen
  \bibfield  {author} {\bibinfo {author} {\bibfnamefont {T.}~\bibnamefont
  {Zhou}}, \bibinfo {author} {\bibfnamefont {L.}~\bibnamefont {L{\"u}}}, \ and\
  \bibinfo {author} {\bibfnamefont {Y.~C.}\ \bibnamefont {Zhang}},\ }\href@noop
  {} {\bibfield  {journal} {\bibinfo  {journal} {Eur. Phys. J. B}\ }\textbf
  {\bibinfo {volume} {71}},\ \bibinfo {pages} {623} (\bibinfo {year}
  {2009})}\BibitemShut {NoStop}%
\bibitem [{\citenamefont {L{\"u}}\ \emph {et~al.}(2009)\citenamefont {L{\"u}},
  \citenamefont {Jin},\ and\ \citenamefont {Zhou}}]{lu2009similarity}%
  \BibitemOpen
  \bibfield  {author} {\bibinfo {author} {\bibfnamefont {L.}~\bibnamefont
  {L{\"u}}}, \bibinfo {author} {\bibfnamefont {C.-H.}\ \bibnamefont {Jin}}, \
  and\ \bibinfo {author} {\bibfnamefont {T.}~\bibnamefont {Zhou}},\ }\href@noop
  {} {\bibfield  {journal} {\bibinfo  {journal} {Phys. Rev. E}\ }\textbf
  {\bibinfo {volume} {80}},\ \bibinfo {pages} {046122} (\bibinfo {year}
  {2009})}\BibitemShut {NoStop}%
\bibitem [{\citenamefont {Papadimitriou}\ \emph {et~al.}(2012)\citenamefont
  {Papadimitriou}, \citenamefont {Symeonidis},\ and\ \citenamefont
  {Manolopoulos}}]{papadimitriou2012fast}%
  \BibitemOpen
  \bibfield  {author} {\bibinfo {author} {\bibfnamefont {A.}~\bibnamefont
  {Papadimitriou}}, \bibinfo {author} {\bibfnamefont {P.}~\bibnamefont
  {Symeonidis}}, \ and\ \bibinfo {author} {\bibfnamefont {Y.}~\bibnamefont
  {Manolopoulos}},\ }\href@noop {} {\bibfield  {journal} {\bibinfo  {journal}
  {J. Syst. Softw.}\ }\textbf {\bibinfo {volume} {85}},\ \bibinfo {pages}
  {2119} (\bibinfo {year} {2012})}\BibitemShut {NoStop}%
\bibitem [{\citenamefont {Liu}\ and\ \citenamefont
  {L{\"u}}(2010)}]{liu2010link}%
  \BibitemOpen
  \bibfield  {author} {\bibinfo {author} {\bibfnamefont {W.}~\bibnamefont
  {Liu}}\ and\ \bibinfo {author} {\bibfnamefont {L.}~\bibnamefont {L{\"u}}},\
  }\href@noop {} {\bibfield  {journal} {\bibinfo  {journal} {EPL}\ }\textbf
  {\bibinfo {volume} {89}},\ \bibinfo {pages} {58007} (\bibinfo {year}
  {2010})}\BibitemShut {NoStop}%
\bibitem [{\citenamefont {Goldberg}\ \emph {et~al.}(1992)\citenamefont
  {Goldberg}, \citenamefont {Nichols}, \citenamefont {Oki},\ and\ \citenamefont
  {Terry}}]{goldberg1992using}%
  \BibitemOpen
  \bibfield  {author} {\bibinfo {author} {\bibfnamefont {D.}~\bibnamefont
  {Goldberg}}, \bibinfo {author} {\bibfnamefont {D.}~\bibnamefont {Nichols}},
  \bibinfo {author} {\bibfnamefont {B.~M.}\ \bibnamefont {Oki}}, \ and\
  \bibinfo {author} {\bibfnamefont {D.}~\bibnamefont {Terry}},\ }\href@noop {}
  {\bibfield  {journal} {\bibinfo  {journal} {Comm. ACM}\ }\textbf {\bibinfo
  {volume} {35}},\ \bibinfo {pages} {61} (\bibinfo {year} {1992})}\BibitemShut
  {NoStop}%
\bibitem [{\citenamefont {Hanley}\ and\ \citenamefont
  {McNeil}(1983)}]{hanley1983method}%
  \BibitemOpen
  \bibfield  {author} {\bibinfo {author} {\bibfnamefont {J.~A.}\ \bibnamefont
  {Hanley}}\ and\ \bibinfo {author} {\bibfnamefont {B.~J.}\ \bibnamefont
  {McNeil}},\ }\href@noop {} {\bibfield  {journal} {\bibinfo  {journal}
  {Radiology}\ }\textbf {\bibinfo {volume} {148}},\ \bibinfo {pages} {839}
  (\bibinfo {year} {1983})}\BibitemShut {NoStop}%
\bibitem [{\citenamefont {Ou}\ \emph {et~al.}(2007)\citenamefont {Ou},
  \citenamefont {Jin}, \citenamefont {Zhou}, \citenamefont {Wang},\ and\
  \citenamefont {Yin}}]{ou2007power}%
  \BibitemOpen
  \bibfield  {author} {\bibinfo {author} {\bibfnamefont {Q.}~\bibnamefont
  {Ou}}, \bibinfo {author} {\bibfnamefont {Y.-D.}\ \bibnamefont {Jin}},
  \bibinfo {author} {\bibfnamefont {T.}~\bibnamefont {Zhou}}, \bibinfo {author}
  {\bibfnamefont {B.-H.}\ \bibnamefont {Wang}}, \ and\ \bibinfo {author}
  {\bibfnamefont {B.-Q.}\ \bibnamefont {Yin}},\ }\href@noop {} {\bibfield
  {journal} {\bibinfo  {journal} {Phys. Rev. E}\ }\textbf {\bibinfo {volume}
  {75}},\ \bibinfo {pages} {021102} (\bibinfo {year} {2007})}\BibitemShut
  {NoStop}%
\bibitem [{\citenamefont {Yan}\ \emph {et~al.}(2006)\citenamefont {Yan},
  \citenamefont {Zhou}, \citenamefont {Hu}, \citenamefont {Fu},\ and\
  \citenamefont {Wang}}]{yan2006efficient}%
  \BibitemOpen
  \bibfield  {author} {\bibinfo {author} {\bibfnamefont {G.}~\bibnamefont
  {Yan}}, \bibinfo {author} {\bibfnamefont {T.}~\bibnamefont {Zhou}}, \bibinfo
  {author} {\bibfnamefont {B.}~\bibnamefont {Hu}}, \bibinfo {author}
  {\bibfnamefont {Z.-Q.}\ \bibnamefont {Fu}}, \ and\ \bibinfo {author}
  {\bibfnamefont {B.-H.}\ \bibnamefont {Wang}},\ }\href@noop {} {\bibfield
  {journal} {\bibinfo  {journal} {Phys. Rev. E}\ }\textbf {\bibinfo {volume}
  {73}},\ \bibinfo {pages} {046108} (\bibinfo {year} {2006})}\BibitemShut
  {NoStop}%
\bibitem [{\citenamefont {Watts}\ and\ \citenamefont
  {Strogatz}(1998)}]{watts1998collective}%
  \BibitemOpen
  \bibfield  {author} {\bibinfo {author} {\bibfnamefont {D.~J.}\ \bibnamefont
  {Watts}}\ and\ \bibinfo {author} {\bibfnamefont {S.~H.}\ \bibnamefont
  {Strogatz}},\ }\href@noop {} {\bibfield  {journal} {\bibinfo  {journal}
  {Nature}\ }\textbf {\bibinfo {volume} {393}},\ \bibinfo {pages} {440}
  (\bibinfo {year} {1998})}\BibitemShut {NoStop}%
\bibitem [{\citenamefont {Newman}(2002)}]{newman2002assortative}%
  \BibitemOpen
  \bibfield  {author} {\bibinfo {author} {\bibfnamefont {M.~E.~J.}\
  \bibnamefont {Newman}},\ }\href@noop {} {\bibfield  {journal} {\bibinfo
  {journal} {Phys. Rev. Lett.}\ }\textbf {\bibinfo {volume} {89}},\ \bibinfo
  {pages} {208701} (\bibinfo {year} {2002})}\BibitemShut {NoStop}%
\bibitem [{Note1()}]{Note1}%
  \BibitemOpen
  \bibinfo {note} {Some data sets are freely downloaded from the following
  academic web sites: http://vlado.fmf.uni-lj.si/pub/networks/data,
  http://wiki.gephi.org/index.php?title=Datasets,
  http://lovro.lpt.fri.uni-lj.si/support.jsp,
  http://konect.uni-koblenz.de/networks/, and
  http://www.linkprediction.org/index.php/link/resource/data.}\BibitemShut
  {Stop}%
\bibitem [{\citenamefont {Batagelj}\ and\ \citenamefont
  {Mrvar}(1998)}]{batagelj1998pajek}%
  \BibitemOpen
  \bibfield  {author} {\bibinfo {author} {\bibfnamefont {V.}~\bibnamefont
  {Batagelj}}\ and\ \bibinfo {author} {\bibfnamefont {A.}~\bibnamefont
  {Mrvar}},\ }\href@noop {} {\bibfield  {journal} {\bibinfo  {journal}
  {Connections}\ }\textbf {\bibinfo {volume} {21}},\ \bibinfo {pages} {47}
  (\bibinfo {year} {1998})}\BibitemShut {NoStop}%
\bibitem [{\citenamefont {Bu}\ \emph {et~al.}(2003)\citenamefont {Bu},
  \citenamefont {Zhao}, \citenamefont {Cai}, \citenamefont {Xue}, \citenamefont
  {Zhu}, \citenamefont {Lu}, \citenamefont {Zhang}, \citenamefont {Sun},
  \citenamefont {Ling}, \citenamefont {Zhang}, \citenamefont {Li},\ and\
  \citenamefont {Chen}}]{Yeast}%
  \BibitemOpen
  \bibfield  {author} {\bibinfo {author} {\bibfnamefont {D.}~\bibnamefont
  {Bu}}, \bibinfo {author} {\bibfnamefont {Y.}~\bibnamefont {Zhao}}, \bibinfo
  {author} {\bibfnamefont {L.}~\bibnamefont {Cai}}, \bibinfo {author}
  {\bibfnamefont {H.}~\bibnamefont {Xue}}, \bibinfo {author} {\bibfnamefont
  {X.}~\bibnamefont {Zhu}}, \bibinfo {author} {\bibfnamefont {H.}~\bibnamefont
  {Lu}}, \bibinfo {author} {\bibfnamefont {J.}~\bibnamefont {Zhang}}, \bibinfo
  {author} {\bibfnamefont {S.}~\bibnamefont {Sun}}, \bibinfo {author}
  {\bibfnamefont {L.}~\bibnamefont {Ling}}, \bibinfo {author} {\bibfnamefont
  {N.}~\bibnamefont {Zhang}}, \bibinfo {author} {\bibfnamefont
  {G.}~\bibnamefont {Li}}, \ and\ \bibinfo {author} {\bibfnamefont
  {R.}~\bibnamefont {Chen}},\ }\href@noop {} {\bibfield  {journal} {\bibinfo
  {journal} {Nucleic Acids Res.}\ }\textbf {\bibinfo {volume} {31}},\ \bibinfo
  {pages} {2443} (\bibinfo {year} {2003})}\BibitemShut {NoStop}%
\bibitem [{\citenamefont {Newman}(2006)}]{newman2006finding}%
  \BibitemOpen
  \bibfield  {author} {\bibinfo {author} {\bibfnamefont {M.~E.~J.}\
  \bibnamefont {Newman}},\ }\href@noop {} {\bibfield  {journal} {\bibinfo
  {journal} {Phys. Rev. E}\ }\textbf {\bibinfo {volume} {74}},\ \bibinfo
  {pages} {036104} (\bibinfo {year} {2006})}\BibitemShut {NoStop}%
\bibitem [{\citenamefont {Gleiser}\ and\ \citenamefont
  {Danon}(2003)}]{gleiser2003community}%
  \BibitemOpen
  \bibfield  {author} {\bibinfo {author} {\bibfnamefont {P.~M.}\ \bibnamefont
  {Gleiser}}\ and\ \bibinfo {author} {\bibfnamefont {L.}~\bibnamefont
  {Danon}},\ }\href@noop {} {\bibfield  {journal} {\bibinfo  {journal} {Adv.
  Complex Syst.}\ }\textbf {\bibinfo {volume} {6}},\ \bibinfo {pages} {565}
  (\bibinfo {year} {2003})}\BibitemShut {NoStop}%
\bibitem [{\citenamefont {Blagus}\ \emph {et~al.}(2012)\citenamefont {Blagus},
  \citenamefont {{\v{S}}ubelj},\ and\ \citenamefont {Bajec}}]{blagus2012self}%
  \BibitemOpen
  \bibfield  {author} {\bibinfo {author} {\bibfnamefont {N.}~\bibnamefont
  {Blagus}}, \bibinfo {author} {\bibfnamefont {L.}~\bibnamefont
  {{\v{S}}ubelj}}, \ and\ \bibinfo {author} {\bibfnamefont {M.}~\bibnamefont
  {Bajec}},\ }\href@noop {} {\bibfield  {journal} {\bibinfo  {journal} {Physica
  A}\ }\textbf {\bibinfo {volume} {391}},\ \bibinfo {pages} {2794} (\bibinfo
  {year} {2012})}\BibitemShut {NoStop}%
\bibitem [{\citenamefont {Guimera}\ \emph {et~al.}(2003)\citenamefont
  {Guimera}, \citenamefont {Danon}, \citenamefont {Diaz-Guilera}, \citenamefont
  {Giralt},\ and\ \citenamefont {Arenas}}]{guimera2003self}%
  \BibitemOpen
  \bibfield  {author} {\bibinfo {author} {\bibfnamefont {R.}~\bibnamefont
  {Guimera}}, \bibinfo {author} {\bibfnamefont {L.}~\bibnamefont {Danon}},
  \bibinfo {author} {\bibfnamefont {A.}~\bibnamefont {Diaz-Guilera}}, \bibinfo
  {author} {\bibfnamefont {F.}~\bibnamefont {Giralt}}, \ and\ \bibinfo {author}
  {\bibfnamefont {A.}~\bibnamefont {Arenas}},\ }\href@noop {} {\bibfield
  {journal} {\bibinfo  {journal} {Phys. Rev. E}\ }\textbf {\bibinfo {volume}
  {68}},\ \bibinfo {pages} {065103} (\bibinfo {year} {2003})}\BibitemShut
  {NoStop}%
\bibitem [{\citenamefont {Isella}\ \emph {et~al.}(2011)\citenamefont {Isella},
  \citenamefont {Stehl{\'e}}, \citenamefont {Barrat}, \citenamefont {Cattuto},
  \citenamefont {Pinton},\ and\ \citenamefont {Van~den Broeck}}]{isella2011s}%
  \BibitemOpen
  \bibfield  {author} {\bibinfo {author} {\bibfnamefont {L.}~\bibnamefont
  {Isella}}, \bibinfo {author} {\bibfnamefont {J.}~\bibnamefont {Stehl{\'e}}},
  \bibinfo {author} {\bibfnamefont {A.}~\bibnamefont {Barrat}}, \bibinfo
  {author} {\bibfnamefont {C.}~\bibnamefont {Cattuto}}, \bibinfo {author}
  {\bibfnamefont {J.-F.}\ \bibnamefont {Pinton}}, \ and\ \bibinfo {author}
  {\bibfnamefont {W.}~\bibnamefont {Van~den Broeck}},\ }\href@noop {}
  {\bibfield  {journal} {\bibinfo  {journal} {J. Theor. Biol.}\ }\textbf
  {\bibinfo {volume} {271}},\ \bibinfo {pages} {166} (\bibinfo {year}
  {2011})}\BibitemShut {NoStop}%
\bibitem [{\citenamefont {Van~Welden}(2004)}]{van2004mapping}%
  \BibitemOpen
  \bibfield  {author} {\bibinfo {author} {\bibfnamefont {D.}~\bibnamefont
  {Van~Welden}},\ }in\ \href@noop {} {\emph {\bibinfo {booktitle}
  {FUBUTEC'2004: 1st future business technology conference}}}\ (\bibinfo
  {organization} {EUROSIS},\ \bibinfo {year} {2004})\ pp.\ \bibinfo {pages}
  {55--59}\BibitemShut {NoStop}%
\bibitem [{\citenamefont {Opsahl}\ and\ \citenamefont
  {Panzarasa}(2009)}]{opsahl2009clustering}%
  \BibitemOpen
  \bibfield  {author} {\bibinfo {author} {\bibfnamefont {T.}~\bibnamefont
  {Opsahl}}\ and\ \bibinfo {author} {\bibfnamefont {P.}~\bibnamefont
  {Panzarasa}},\ }\href@noop {} {\bibfield  {journal} {\bibinfo  {journal}
  {Soc. Netw.}\ }\textbf {\bibinfo {volume} {31}},\ \bibinfo {pages} {155}
  (\bibinfo {year} {2009})}\BibitemShut {NoStop}%
\bibitem [{\citenamefont {Bascompte}(2004)}]{bascompte2004food}%
  \BibitemOpen
  \bibfield  {author} {\bibinfo {author} {\bibfnamefont {J.}~\bibnamefont
  {Bascompte}},\ }\href@noop {} {\bibfield  {journal} {\bibinfo  {journal}
  {Ecology}\ }\textbf {\bibinfo {volume} {85}},\ \bibinfo {pages} {352}
  (\bibinfo {year} {2004})}\BibitemShut {NoStop}%
\bibitem [{\citenamefont {Hummon}\ and\ \citenamefont
  {Dereian}(1989)}]{hummon1989connectivity}%
  \BibitemOpen
  \bibfield  {author} {\bibinfo {author} {\bibfnamefont {N.~P.}\ \bibnamefont
  {Hummon}}\ and\ \bibinfo {author} {\bibfnamefont {P.}~\bibnamefont
  {Dereian}},\ }\href@noop {} {\bibfield  {journal} {\bibinfo  {journal} {Soci.
  Netw.}\ }\textbf {\bibinfo {volume} {11}},\ \bibinfo {pages} {39} (\bibinfo
  {year} {1989})}\BibitemShut {NoStop}%
\bibitem [{\citenamefont {Guimera}\ \emph {et~al.}(2005)\citenamefont
  {Guimera}, \citenamefont {Mossa}, \citenamefont {Turtschi},\ and\
  \citenamefont {Amaral}}]{guimera2005worldwide}%
  \BibitemOpen
  \bibfield  {author} {\bibinfo {author} {\bibfnamefont {R.}~\bibnamefont
  {Guimera}}, \bibinfo {author} {\bibfnamefont {S.}~\bibnamefont {Mossa}},
  \bibinfo {author} {\bibfnamefont {A.}~\bibnamefont {Turtschi}}, \ and\
  \bibinfo {author} {\bibfnamefont {L.~N.}\ \bibnamefont {Amaral}},\
  }\href@noop {} {\bibfield  {journal} {\bibinfo  {journal} {Proc. Natl. Acad.
  Sci. USA}\ }\textbf {\bibinfo {volume} {102}},\ \bibinfo {pages} {7794}
  (\bibinfo {year} {2005})}\BibitemShut {NoStop}%
\end{thebibliography}%

\end{document}